\documentclass[10pt,conference]{IEEEtran} 

\usepackage[utf8]{inputenc}
\usepackage[T1]{fontenc}
\usepackage{microtype}
\usepackage{graphicx}
\usepackage{nccmath}
\usepackage{xcolor}
\usepackage{newtxmath}
\usepackage{mathtools}
\usepackage{hyperref}
\usepackage{float}
\usepackage{cite}
\usepackage[caption=false,font=footnotesize]{subfig}
\usepackage[textsize=scriptsize]{todonotes}

\newcommand{\email}[1]{\href{mailto:#1}{\texttt{#1}}}

\newcommand{\EV}[1]{\vmathbb{E}[#1]}

\newcommand{\phm}{\phantom-}

\begin{document}

\title{A Linear Algebraic Framework \\
  for Quantum Internet Dynamic Scheduling}
\author{\IEEEauthorblockN{\hfill Paolo Fittipaldi\IEEEauthorrefmark{1} \hfill
  Anastasios Giovanidis\IEEEauthorrefmark{2} \hfill  
  Frédéric Grosshans\IEEEauthorrefmark{3} \hfill}
\IEEEauthorblockA{Sorbonne Université,  CNRS, LIP6, F-75005 Paris, France \\ 
 \hfill
 \IEEEauthorrefmark{1}\email{paolo.fittipaldi@lip6.fr}
 \hfill \IEEEauthorrefmark{2}\email{anastasios.giovanidis@lip6.fr} 
 \hfill 
 \IEEEauthorrefmark{3}\email{frederic.grosshans@lip6.fr}
 \hfill}}

\maketitle

\begin{abstract}
Future quantum internet aims to enable quantum communication between arbitrary pairs of distant nodes through the sharing of end-to-end entanglement, a universal resource for many quantum applications.
As in classical networks, quantum networks also have to resolve problems related to routing and satisfaction of service at a sufficient rate.
We deal here with the problem of scheduling when multiple commodities must be served through a quantum network based on first generation quantum repeaters, or quantum switches.
To this end, we introduce a novel discrete-time algebraic model for arbitrary network topology, including transmission and memory losses, and adapted to dynamic scheduling decisions.

Our algebraic model allows the scheduler to use the storage of temporary intermediate links to optimize the performance, depending on the information availability, ranging from full global information for a centralized scheduler to partial local information for a distributed one.
 As an illustrative example, we compare a simple greedy scheduling policy with several Max-Weight inspired scheduling policies and illustrate the resulting achievable rate regions for two competing pairs of clients through a network.
\end{abstract}

\section{Introduction}

Designing the Quantum Internet raises many challenges to network scientists and quantum physicists alike. Among them, the one of designing a scheduling policy is particularly familiar to the former: whenever multiple users wish to communicate through a classical network, a scheduling policy regulates fair and efficient relay of the data packets inside queues across the network routes. In the context of networks based on first generation quantum repeaters\cite{1g2g3g} or quantum switches\cite{TowsleySwitchProtocols,TowsleySwitchStochastic}, users are linked through quantum entanglement swapping: letting $AB$ and $BC$ be two links that share one entangled pair each, entanglement swapping allows to teleport their entanglement to the link $AC$ through a local measurement at node $B$. This consumes the $AB$ and $BC$ pairs and yields one $AC$ pair. 
Entangled pairs of qubits, or \emph{ebits}, are a universal resource for quantum communication 
\cite{wilde}: together with classical communications, they allow to implement a wide array of quantum protocols\cite{protocolzoo}. The task of a quantum network is therefore to distribute entanglement to multiple user pairs through a graph of interconnected quantum repeaters. Since the routes of service might not be disjoint, distribution along routes that share a subpath creates conflict that must be carefully mediated by the scheduler. In practice, a quantum scheduler must determine which swapping operations to perform at a given time, balancing between serving user requests and keeping a margin to improve future performance. Notice how the scheduling challenge naturally descends from the introduction of 
memories: without them,
large scale networks only have the option to either swap pairs as soon as they are distributed or waste them. This work actively exploits the memory in the scheduling process, which yields the possibility for the network control system to store some pairs to swap at a following time, leveraging this additional degree of freedom through carefully taken scheduling decisions in a similar way to what \cite{RoutingMemoriesWehnerKerenidis} shows for routing.
Throughout this work, the routing in the network is assumed to be fixed and known, i.e., every user pair comes with a set of routes along which the network will distribute entanglement to serve requests.
We propose a linear algebraic discrete time model for quantum scheduling including all the previously mentioned factors which is suitable for any network topology including heterogeneous ones, and is dynamically 
controlled, i.e., the scheduling decisions are taken in real time given some degree of information on the current network state. The algebraic model is presented and formalized, and then applied to different scheduling policies: a greedy one where each node swaps randomly as soon as there is pair availability, a Max Weight\cite{Tassiulas} inspired one that has full information about the state of the network at any time and can therefore provide a best case scenario, and an intermediate one where each node solves its own individual Max Weight problem, using only local information. The achievable rate regions of these policies are shown and compared. The provided comparison ranges also in terms of localization: a global scheduler is implemented as a central block outside the network that receives information about the system's state and broadcasts back a decision, while a localized one gives the nodes themselves the authority to decide. As will be analyzed in the dedicated section, varying the degree of localization of the scheduler changes the amount of time required for classical communication, indirectly affecting the achievable performance of the scheduler.
The rest of this work has the following organization: Section \ref{sec_context} reviews the scientific context around this work, Section \ref{sec_sysdesc} describes in detail the system we are modeling and Section \ref{sec_application} shows application of the algebraic model to real scheduling policies. Section \ref{sec_numresults} shows the numerical results we obtained through our model, and Section \ref{sec_conclusion} concludes the paper. \section{Context and relevance of this work}
\label{sec_context}
A large amount of work in the quantum internet field consists in adapting classical network theory concepts through novel ideas that bridge the gap. An introduction to the subject and definition of the quantum network stack can be found in \cite{QStack}. Delving deeper, one may see this work as an extension of \cite{TowsleyGrid}, which treats the problem of routing without scheduling: this work deals with scheduling and adds a treatment of memory and loss.  In \cite{TowsleySwitchStochastic,TowsleySwitchProtocols}, a full stochastic analysis of a single quantum switch is provided and some scheduling policies are implemented on it: we state a similar problem but on an arbitrary network topology, deriving results that should prove relevant on several network scales. \cite{vasantham} examines the application of a Max Weight policy to quantum networks, akin to the last section of this work, the difference being that our effort focuses on the general scheduling model, with the Max Weight policy being provided as an example of its applicability. Finally, an optimal theoretical bound for entanglement distribution across a network with a single commodity is derived in \cite{DaiScheduling} and expanded upon in \cite{TowsleyScheduling}. Our work extends the treatment to multiple commodities on any arbitrary topology accounting for degrees of freedom such as memory, ebit generation statistics, technology imperfections (e.g., memory and fiber losses) and scheduling policy.

An important contribution of this work is the novel application of an idea similar to \cite{RoutingMemoriesWehnerKerenidis,TowsleyScheduling}, in a non-trivial way to general topologies and multiple commodities: the introduction of memory at the nodes allows them to decide between employing entangled pairs for swapping or keeping them for future use. The deeper implication of this point is that the network is free to create intermediate links and store them: this leads to distributing pairs across a service route in a ``growing'' fashion, that both increases performance and removes the need for end-to-end link state information.

\section{System Description}
\label{sec_sysdesc}
We use the following notation convention: lower case for scalars ($x$), bold lower case for vectors ($\mathbf{x}$), bold upper case for matrices ($\mathbf{X}$) and calligraphic upper case for sets ($\mathcal{X}$). Well-known matrices such as the identity matrix or the null matrix are indicated in blackboard bold and subscripted with their dimension ($\vmathbb{I}_{n}$, $\vmathbb{0}_{n\times m}$).

The physical system that will be modeled is a network of quantum switches. These are devices that can hold qubits and perform entanglement swapping across multiple pairs of clients, akin to a quantum repeater with multiple possible linking paths. Given an arbitrary connected graph $\mathcal{G} = (\mathcal{V},\mathcal{E})$, the switches are deployed at the locations specified by the vertices of $\mathcal{G}$ and interconnected by lossy fiber links running along each edge $(i,j) \in \mathcal{E}$. 
Every switch has a number of memory slots, assumed to be infinite in this work, in which qubits may be stored. Ebits (pairs of entangled qubits) are generated by each fiber link with a given constant average rate, which may be different for each link, and stored inside memories at the end nodes of the respective link. Among the network nodes, 
there are $n$ fixed pairs
$\{(\mathit{Alice}_1, \mathit{Bob}_1),\ldots,(\mathit{Alice}_n, \mathit{Bob}_n)\}$
that request ebits in a random way to realize a generic application. The $(\mathit{Alice}_n, \mathit{Bob}_n)$ pairs are connected by fixed known routes that are not necessarily disjoint and therefore can create congestion across some of the network links, that needs to be managed by a scheduler.
The task of the network is to distribute ebits to serve demands through entanglement swapping, while being hindered by loss: other than the losses across the fiber links, an additional form of loss is 
tied to memory imperfections, which cause stored qubits to effectively disappear and entanglement to be lost.
Memory and fiber losses are the only two sources of imperfection that are accounted for in this paper: swaps in the switches are assumed to always succeed and memory slots at each switch are infinite, but neither of these assumptions is too limiting and they could easily be lifted in following research. 
For simplicity reasons, we neglect  sources of state degradation other than losses in this introduction of our algebraic model, since they require a more detailed description of the quantum state of the ebits, and lead to more complex multiobjective routing and scheduling algorithms \cite{bugalho,routingkaushik}. Note also that our model also applies directly to more long-term quantum networks, where the ebits are logical error-corrected ebits. 

For practical reasons, our model assumes a discrete time: swapping operations are supposed to occur at fixed time intervals, thus it is natural to take a discrete time step $\Delta t$ as the time unit of interest. 
Between two subsequent clock ticks, the system is free to evolve, and at the end of each time step a scheduling decision is taken. This places a lower bound on $\Delta t$: no decision can happen before all information has been successfully communicated to all deciding agents, thus $\Delta t$ must be at least as large as the classical communication delay introduced by state-related information exchange. This in turn introduces a tradeoff: large $\Delta t$ means that information can travel farther before the scheduling decision is taken (allowing for larger networks or scheduling policies that require several physically spaced nodes to communicate), but it increases losses, as detailed below, and it introduces the issue of stale information: during the time it takes for state information to reach its recipient, the system continues its stochastic evolution, making the communicated data less relevant by the time it reaches the place where the scheduling decision is taken.
To model ebits stored at memory nodes, the concept of an \emph{ebit queue} is introduced: each pair of nodes $e = (i,j)$ inside the extended edge set $\tilde{\mathcal{E}}=\mathcal{V}\times\mathcal{V}$ is said to possess an ebit queue $q_{ij}(t)$. Furthermore, among ebit queues, every $q_{ij}(t)$ associated to an edge $(i,j)\in\mathcal{E}$ equipped with fiber is called a \emph{physical queue}, while all other ebit queues are called \emph{virtual queues}. Ebit queues are an abstraction for memory slots on pairs of nodes: generation, loss and swapping may be modeled as addition, subtraction and exchange along the relevant queues. 
At each time step, every fiber link --- and thus every physical queue --- generates a number of 
ebits $a_{ij}(t)$. This can model different ebit generation processes: either the physical link $(i,j)$ corresponds to a twin photon 
 source which propagates to the switches $i$ or $j$, or alternatively it corresponds to two photons --- each one entangled with one of the switches --- 
 that meet in the middle of the link where they are subjected to a Bell state measurement (BSM). It can also model more elaborate entanglement distillation protocols or error correction based protocols based on logical qubits. Since most of these processes are probabilistic in
 nature  $a_{ij}(t)$ will here be treated as a random process, which we assume to be Poissonian of mean value $\alpha_{ij}$, constant in time. 
 This allows to model the link imperfections --- finite brightness of the source, propagation losses, finite
 success probability of photonic BSMs, etc.\@ --- as a Poisson filtration process, which simply decreases the value of $\alpha_{ij}$. Note that $a_{ij}(t)$ need not be Poissonian: more elaborate models for $a_{ij}(t)$ are possible, e.g.\@ to model an almost perfect pulsed periodic twin-photon source. 
The modelization of memory loss is slightly more complex: a qubit stored inside a quantum memory has a probability $\eta$ to survive for a time $\Delta t$ that is exponentially decreasing as 
$\eta = \exp{\left(-\frac{\Delta t}{\tau}\right)}$, 
 where $\tau$ is the expected lifetime of a qubit in the memory, a technological parameter expected to vary between nanoseconds and milliseconds in near term implementations \cite{LKBMem}.
By setting $\Delta t$ to the duration of a time step, we obtain the probability $\eta$ for a stored ebit to survive for one time step. The number of timesteps an ebit survives in memory is then given by the geometric distribution
 defined by $\eta$. It is easy to show its mean value $\tfrac{1}{1-\eta}$ tends to the expected
 $\tfrac{\tau}{\Delta t}$ for small $\frac{\Delta t}{\tau}$, $\tau$ being the actual lifetime of the memories. 
 The remaining difference is an effect of the dicretization. Looking now at all the ebits in queue $q_{ij}(t)$ collectively from one timestep to the next, their losses are modeled by a binomially distributed random variable $\ell_{ij}(t)$, with as many trials as there are ebits stored in queue $(i,j)$ and probability to lose one pair $1 - \eta$. Accounting for losses in such a time-dependent way makes the presented framework valid also as a tool to determine the optimal frequency at which scheduling decision should be taken, given the technological parameters.

\begin{figure}[t!]
\centering
\includegraphics[width=.9\linewidth]{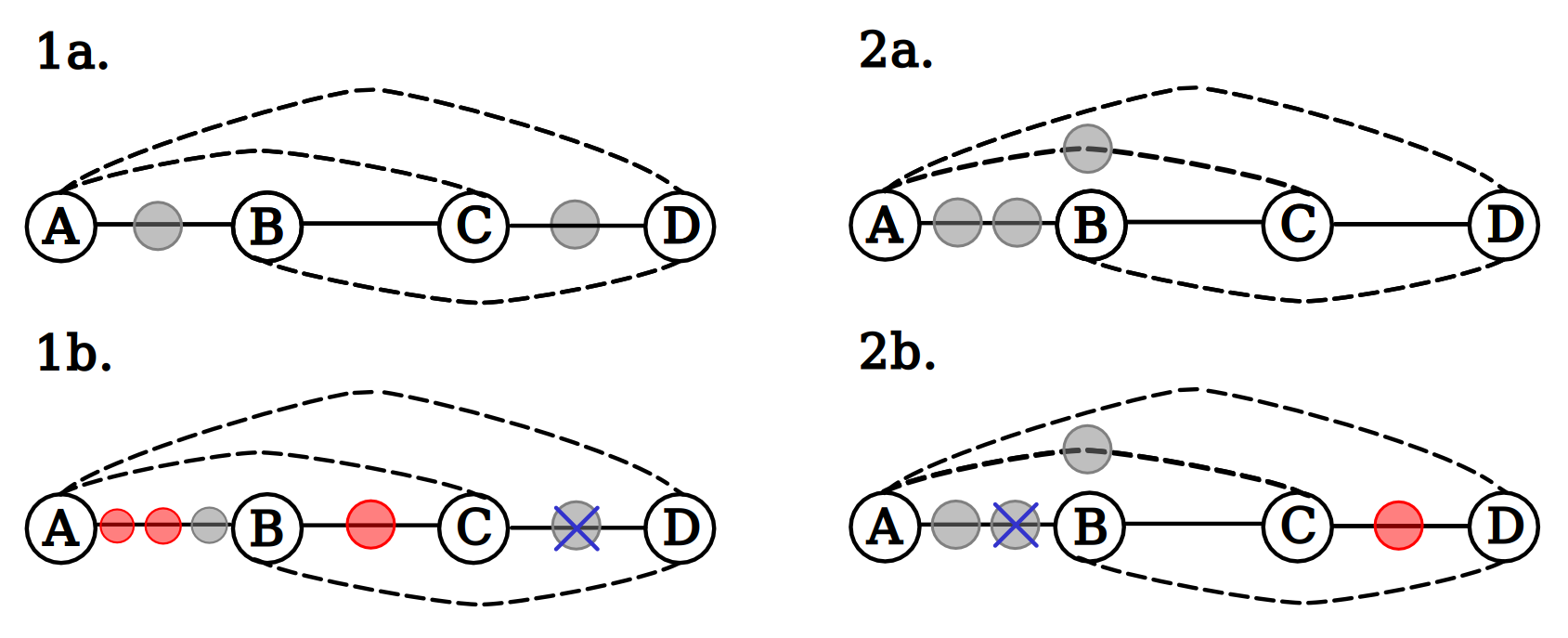}
\caption{Example of simulation of two time steps over the whole topology. Continuous lines represent physical queues and dashed lines virtual ones. Grey circles represent ebits that were in the queue at the beginning of a time step, red ones ebits that arrived during that time step. Blue crosses represent loss of an ebit. Upper figures (a) at the beginning of the time step, lower figures (b) at the end of the time step.} 
\label{fig_simexample}
\end{figure}

\begin{figure}[t!]
\centering
\includegraphics[width=.9\linewidth]{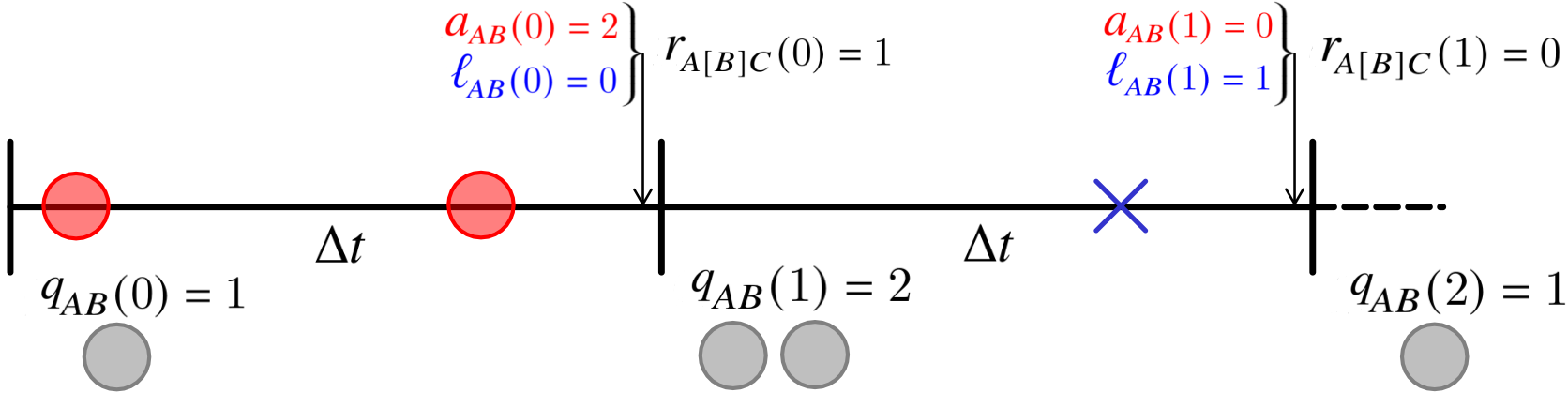}
\caption{Example of two time steps from the point of view of queue $AB$. Circles and crosses have the same meaning as fig.\ref{fig_simexample}. Queue snapshots are taken at the very beginning of a time step, while arrivals and losses happen during but are only assessed at the end of the step as soon as the scheduling decision is taken. Ebits arriving during the current time step are not subject to losses in this model.}
\label{fig_simtiming}
\vspace{-0.5cm}
\end{figure}

For what concerns scheduling decisions, let $r_{i[j]k}(t)$ indicate the number of swapping operations that happen at a given time step, at node $j$, from queues $(i,j)$ and $(j,k)$ to queue $(i,k)$: as a notation example, $r_{A[B]C}(2) = 3$ indicates three swapping operations at node $B$ from queues $AB$, $BC$ to $AC$ at time step $2$. Every node will be associated to as many $r_{i[j]k}(t)$ variables as there are swapping operations that can be performed at the node in the given routing, and the scheduler's task will be to set such variables to control the network's behavior. To clarify, suppose to have the service route $ABCD$ across the users $A$ and $D$, as shown in Fig.~\ref{fig_simexample}. Assume the average arrival rates to be $\alpha_{AB}$, $\alpha_{BC}$ and $\alpha_{CD} = 1\  \text{(time steps)}^{-1}$. Lastly, assume that all the memories in the system have $\eta = 0.9$ storage-and-retrieval efficiency. Fig. \ref{fig_simtiming} shows the same test run but focusing on queue $AB$, to highlight the timing of the simulation.

$\bullet$ During time step 0:
\begin{enumerate}
    \item At the beginning of the time step, the queue states are: $q_{AB}(0) = q_{CD}(0) = 1$, $q_{BC}(0) = 0$
    \item At the end of the time step, new ebits have been generated ($a_{AB}(0) = 2$, $a_{BC}(0) = 1$) and one has been lost ($a_{CD}(0) = 1$). The scheduling decision is taken from this configuration as $r_{A[B]C}(0) = 1$: one swap at node $B$ from queues $AB$ and $BC$ to $AC$.
\end{enumerate}

$\bullet$ During time step 1:
\begin{enumerate}
    \item The initial configuration sees two stored pairs in $AB$ which were not employed in the last time step ($q_{AB}(1) = 2$) and the freshly swapped one in $AC$ ($q_{AC}(1) = 1$).
    \item Throughout the time step, one pair was lost across $AB$ ($\ell_{AB}(1) = 1$) and one generated across $CD$. The scheduler may now decide $r_{A[C]D}(1) = 1$ to move to $AD$ or store the pairs for future use.
\end{enumerate}
In term of ebits, a given transition $i[j]k$ is
\textit{incoming} for queue $(i,k)$ and \emph{outgoing} for queues $(i,j)$ and $(j,k)$. A queue's evolution can therefore be summarized as follows, i.t.\@ being a shorthand 
 for \emph{incoming transitions}, o.t.\@ for \emph{outgoing transitions}:
\begin{multline}
    q_{ij}(t+1) = q_{ij}(t) + a_{ij}(t) - \ell_{ij}(t) \\ - \smashoperator{\sum_{o \in \text{o.t.}}}r_o(t) + \smashoperator{\sum_{k \in \text{i.t.}}}r_k(t). 
    \label{scalarevolQ}
\end{multline}
It should be noted that, while all terms of (\ref{scalarevolQ}) are calculated for every queue, $a_{ij}(t)$ across a virtual queue will always be zero, because virtual queues do not generate ebits. 
Conversely, it should be stressed that the loss term $\ell_{ij}(t)$ is calculated in the same way for all queues, because ebit storage is always handled by memories at the network nodes: the physical/virtual queue is a mere modeling artifact to keep track of which memories store the two parts of a given ebit. 
A description of the whole system requires $|\tilde{\mathcal{E}}|$ equations like (\ref{scalarevolQ}).
To keep things compact, it is useful to define some vector terms. The first ones are $\mathbf{q}(t)$, $\mathbf{a}(t)$ and $\boldsymbol\ell(t)$, whose $N_{\text{queues}}$ entries correspond to the individual $q_{ij}(t)$, $a_{ij}(t)$ and $\ell_{ij}(t)$ values (the ordering is irrelevant as long as it is consistent). Furthermore, since the effect of swapping on the queues is linear, it is possible to describe it by introducing the vector $\mathbf{r}(t)$, 
which has $N_{\text{transitions}}$ elements --- as many as there are allowed transitions 
--- and a  matrix $\mathbf{M}$ with $N_\text{queues}$ rows and $N_\text{transitions}$ columns. 
The $\mathbf{r}(t)$ vector stores all the $r_{i[j]k}(t)$ terms, and embodies the scheduling decision, while the $\mathbf{M}$ matrix introduces an efficient encoding of the network topology and the service routes: For each of its columns, associated to transition $i[j]k$, the $\mathbf{M}$ matrix has $-1$ on the rows associated to queues $(i,j)$ and $(j,k)$, and $+1$ on the row associated to queue $(i,k)$. All other terms are zero. System-wide queue evolution can be restated as the following simple linear equation:
\begin{align}
\label{vectorialevolQ}
\mathbf{q}(t+1)  & = \mathbf{q}(t) - \boldsymbol\ell(t) + \mathbf{a}(t) + \mathbf{Mr}(t).
\end{align}
Notice that eq. (\ref{vectorialevolQ}) entails an implicit assumption that the $\mathbf{r}(t)$ be not only a scheduling decision, but a \emph{feasible} scheduling decision, i.e., one that does not cause the queues to turn negative. The meaning of this assumption will be clear when discussing partially informed scheduling policies.
An example of the $\mathbf{M}$ matrix is given in table \ref{tab:Mexample}.
\begin{table}[tb]
    \centering
    \caption{$\mathbf{M}$ matrix for the linear $ABCD$ network}
    \begin{tabular}{l | c c c c}
    & ${A[B]C}$ & ${B[C]D}$ & ${A[B]D}$ & ${A[C]D}$\\\hline
    $AB$ & $-1$ & $\phm0$ & $-1$ & $\phm0$ \\
    $BC$ & $-1$ & $-1$ & $\phm0$ & $\phm0$ \\ 
    $CD$ & $\phm0$ & $-1$ & $\phm0$ & $-1$ \\
    $AC$ & $+1$ & $\phm0$ & $\phm0$ & $-1$ \\
    $BD$ & $\phm0$ & $+1$ & $-1$ & $\phm0$ \\
    $AD$ & $\phm0$ & $\phm0$ & $+1$ & $+1$ \\
    \end{tabular}
    \label{tab:Mexample}
    \vspace{-0.5cm}
\end{table}
The sum of each column of $\mathbf{M}$ is $-1$: this work only employs binary swaps, and each binary swap consumes two ebits and produces one, balancing to an overall $-1$ on the total ebit count. Given a set of service routes, the $\mathbf{M}$ matrix can be automatically generated with the code provided in the application section.
The final piece of the puzzle for ebit queues is consumption: whenever there is availability of entangled pairs across one of the final $(\mathit{Alice}_n,\mathit{Bob}_n)$ pairs, the scheduler must be able to use the available pairs to serve requests. This is implemented in the model by extending the matrix $\mathbf{M}$ to a new $\tilde{\mathbf{M}} = \left[ \mathbf{M} \middle|
-\vmathbb{I}_{N_{\text{queues}}}  \right]$, and the $\mathbf{r}(t)$  vector to have $N_{\text{transitions}} + N_{\text{queues}}$ components. What this extension achieves is to have a set of transitions that only remove one pair from a given queue, modeling actual consumption of the distributed pair by the users. Due to its construction, the $\mathbf{\tilde{M}}$ matrix gives an efficient and concise description of the network topology and how the clients are connected to each other: it is therefore topology and service dependent, and provides all information about both.
Putting it all together, the vector of ebit queues evolves as:
\begin{align}
\label{vectorialevolQwithMtilde}
\mathbf{q}(t+1)  & = \mathbf{q}(t) - \boldsymbol\ell(t) + \mathbf{a}(t) + \mathbf{\tilde{M}r}(t).
\end{align}
where the feasibility assumption from above is still taken. The introduction of consumption completes modeling of the ebits part of the system. However, there is no representation of user requests yet: in a real quantum network, users would request a given number of ebits to cater to a specific application at random times, and a well designed scheduler needs to take user requests into account when controlling the swapping network. Similarly to ebits, demands arriving to the system and
being held for future service are also modeled through queues: alongside every ebit queue, there exists a demand queue $d_{ij}(t)$ that keeps track of the number of user-issued requests (as introduced in \cite{TowsleySwitchProtocols} for a single switch and generalized in this work for an arbitrary topology).

At each time step, every demand queue $d_{ij}(t)$ receives $b_{ij}(t)$ demands, which for simplicity are modeled as a Poisson process with average value $\beta_{ij}$ (as in the case of ebit generation, more complex models that accurately simulate real user demands may be implemented). To maintain the model's uniformity, all edges belonging to $\tilde{\mathcal{E}}$ have a demand queue, but only the ones that are associated to an $(\mathit{Alice}_n, \mathit{Bob}_n)$ pair have nonzero arrivals. All demand queues that are not associated to an $(\mathit{Alice}_n, \mathit{Bob}_n)$ pair are permanently zero. 

Demand queues have a simpler evolution than ebit ones because demands can only be generated, stored and eventually served: they are neither lost nor swapped. 
To model consumption without swapping, we introduce the matrix $\tilde{\mathbf{N}} =  \left[\vmathbb{0}_{N_{\text{queues}}\times N_{\text{transitions}}} \middle| -\vmathbb{I}_{N_{\text{queues}}}\right]$ as a mean of interfacing with the consumption part of the $\mathbf{r}(t)$ vector. The evolution of demand queues is therefore:
\begin{align}
\label{vectorialevolD}
\mathbf{d}(t+1) &= \mathbf{d}(t) + \mathbf{b}(t) + \mathbf{\tilde{N}r}(t)
\end{align}
Notice that the last $N_\text{queues}$ components of the $\mathbf{r}(t)$ vector regulate both demand and ebit consumption: one demand always consumes one ebit. 

\section{Scheduling Application}
\label{sec_application}
The algebraic framework presented above poses as a tool to gauge the performance and requirements of different scheduling policies. In particular, a key point when discussing scheduling policies is the availability of information: a scheduler working with more information will understandably have much better performance but pose harsher requirements on the classical communication infrastructure that accompanies the quantum system. This creates a clear tradeoff between information requirements and performance, and the remainder of this work is dedicated to exploring it, with the ideal result being a scheduler that performs well while needing as little information as possible and with a preference for decentralized policies because they scale better.
 
We start by presenting a greedy scheduler which utilises minimal strictly local information: the greedy scheduler performs swapping 
whenever there is availability of ebits and without adapting to user demand. Therefore, in a greedy scenario, every node randomly links pairs of queues connected to it, until ebit resources are exhausted. 
In spite of the scheduling being random, the greedy scheduler is still aware of routing: 
on a $ABCD$ linear topology with service routes $ABC$ and $BCD$, none of the schedulers examined in this work will create $AD$ entanglement because it is outside the service routes. Despite its performance being much lower than more refined schedulers, the greedy scheduler has the advantage of not requiring any information about the system's state, thus lifting all the classical communication requirements and being completely decentralized. Its main use is as a baseline benchmark, in that any other scheduler must outperform it by a large enough margin to justify the requirement for additional information and/or communication infrastructures. 
 
To gauge the improvement brought by additional information, we also propose an analysis of two Max Weight policies. The Max Weight protocol is a well-known result of network theory \cite{Tassiulas} \cite{vasantham} that revolves around solving a linear program at each time step. As soon as a queue (either ebit or demand) grows, 
the Max Weight scheduler will try to reduce it; the aim is to guarantee a service tailored to user requests on the demand queue side, and also efficient and fair resource exploitation on the ebit queues side, that does not lead to useless accumulation.

To fully explore the range of improvement that information availability can supply, we start from an ideal case of fully informed Max Weight scheduler: at each time step, the scheduler knows the full state of the network ($\mathbf{q}(t),\mathbf{d}(t),\mathbf{a}(t),\boldsymbol\ell(t),\mathbf{b}(t)$...) and can therefore take the best possible decision. The discrete linear problem is stated as:
\begin{eqnarray}
\label{max-weight-prob}
\begin{tabular}{c l l}
$\min$ & $\mathbf{w}^T(t) \cdot \mathbf{r}(t)$ & [global-MW]\\
s.t. & $\mathbf{r}(t)\in\mathcal{R}(t)$, 
\end{tabular}
\end{eqnarray}
with the weights given by
\begin{eqnarray}
\label{w_full}
\mathbf{w}(t) = \gamma(\mathbf{d}(t) + \mathbf{b}(t))^T\tilde{\mathbf{N}} + (\mathbf{q}(t) - \boldsymbol\ell(t) + \mathbf{a}(t))^T\tilde{\mathbf{M}},
\end{eqnarray}
where $\gamma$ is a tunable parameter that allows to prioritize demand queues or ebit queues in the scheduling calculations. The set $\mathcal{R}(t)$ of all possible scheduling decisions $\mathbf{r}(t)$ at time slot $t$ is defined as:
\begin{eqnarray}
\label{constrFULL}
\mathcal{R}(t) & = & \left\{\mathbf{r}(t)\in\mathbb{N}^d
\ \middle|\ -\mathbf{\tilde{M}r}(t)\leq 
\mathbf{q}(t) - \boldsymbol\ell(t) + \mathbf{a}(t)\ \right.\nonumber\\
& & \hspace{+1.5cm} \left. \& \ -\mathbf{\tilde{N}r}\leq 
   \mathbf{d}(t) + \mathbf{b}(t) \right\}
\end{eqnarray}
with $d=N_{\text{transitions}}+N_{\text{queues}}$
a shorthand for the dimension of $\mathbf{r}(t)$. Note that the scheduling decisions are vectors of natural numbers, each decision consuming some number of ebits.

This full knowledge scenario provides a good performance upper bound for Max Weight schedulers and a large performance margin over the greedy policy, as shown in the numerical evaluation section.
Such a fully informed global scheduler is in practice unrealistic as it requires complete information about the instantaneous system state, right before each scheduling decision. In fact, the full information would require a large classical communication delay $\Delta t$, so the scheduler would be based on stale information.
This motivates to propose a scheduling policy that requires less information, while retaining to a certain level the performance benefits seen in the global-MW case.
The policy should rely on a limited amount of exact quantities in the system's state, together with  suitable assumptions about unknown information in order to carry out a scheduling decision that is close to the optimal one. In particular, we propose a Max Weight inspired scheduling policy that works in a partially localized way:
given a network, we assume that all nodes have access to the network's topology, through knowledge over the matrix $\mathbf{M}$ and know average quantities of interest, such as the intensity of Poisson generation processes $\alpha_{ij}, (i,j) \in \mathcal{E}$, 
the memory efficiency $\eta$ and the average demand arrival rates $\beta_{ij}, (i,j) \in \tilde{\mathcal{E}}$. We furthermore assume that every node knows the exact state of the network at the beginning of each time step, i.e., $\mathbf{q}(t)$ and $\mathbf{d}(t)$. Despite $\mathbf{q}(t)$ and $\mathbf{d}(t)$ being global information, we remind the reader that in our model this information is exact for the beginning of the time step, and is communicated during the time step $\Delta t$, to reach all nodes by the end of the slot, when the scheduling decision is made. Moreover, each node has complete information at the end of the slot about all the queues (both physical and virtual) that are directly connected to it. This last piece of information is also available to the greedy scheduler and works as a leverage point that provides great performance improvements.

Each node states its own Max Weight problem for the whole network, combining its own perfect local information with stale information received and with expected values for the other nodes. Denoting as $\mathcal{C}^i$ the set of all edges $e=(i,j) \in \tilde{\mathcal{E}}$ that are connected to node $i$, the information available on node $i$ at step $t$ is
\begin{eqnarray}
\label{info}
\mathcal{I}^i(t) = \left\{\mathbf{q}(t),\mathbf{d}(t),\eta,\beta,\alpha,a_{e}(t),\ell_{e}(t),b_{e}(t),\ \forall e\in\mathcal{C}^i\right\}.\nonumber\\
\end{eqnarray}
By definition $ \forall i,j,t$, $\alpha_{ij} = \EV{a_{ij}(t)}$ and $= \beta_{ij}=\EV{b_{ij}(t)}$, so the vectors of all mean arrivals are $\alpha$ and $\beta$ respectively. The vectors $\mathbf{q}(t),\mathbf{d}(t)$ communicated to node $i$ as well as the mean vectors $\alpha,\beta$ are all of size $\mathit{N_{queues}}$. Moreover, since $\eta$ is the memory efficiency, $\EV{\ell_{ij}(t)} = (1 - \eta)q_{ij}(t)\ \forall i,j,t$.

The linear discrete scheduling problem localised on node $i$ can be stated as:
\begin{eqnarray}
\begin{tabular}{c l l}
$\min$ & $(\mathbf{w}^{i})^T(t) \cdot \mathbf{r}(t)$ & [$i$-local-MW]\\
s.t. & $\mathbf{r}(t)\in\mathcal{R}^{i}(t)$, 
\end{tabular}
\end{eqnarray}
where the weights are given by
\begin{eqnarray}
\label{w}
\mathbf{w}^{i}(t) & = & \gamma\EV{\mathbf{d}(t) + \mathbf{b}(t)|\mathcal{I}^i(t)}^T\tilde{\mathbf{N}} \nonumber\\ 
& + & \EV{\mathbf{q}(t) - \boldsymbol\ell(t) + \mathbf{a}(t)|\mathcal{I}^i(t)}^T\tilde{\mathbf{M}}.
\end{eqnarray}
Where $\gamma$ serves the same purpose as before. The set $\mathcal{R}^{i}(t)$ of all possible scheduling decisions $\mathbf{r}(t)$ at time slot $t$ localised at node $i$ is defined as:
\begin{eqnarray}
\label{constrLOC}
\mathcal{R}^{i}(t) & = & \left\{\mathbf{r}(t)\in\mathbb{N}^d
\ \middle|\ -\tilde{\mathbf{M}}\mathbf{r}\leq 
\EV{\mathbf{q}(t) - \boldsymbol\ell(t) + \mathbf{a}(t)| \mathcal{I}^i(t)}\ \right. \nonumber\\ 
& & \left. \hspace{+1.5cm} \&\ -\tilde{\mathbf{N}}\mathbf{r}(t)\leq 
   \EV{\mathbf{d}(t) + \mathbf{b}(t)|\mathcal{I}^i(t)}\right\}.
\end{eqnarray}
\begin{figure*}
\centering
\subfloat[]{\includegraphics[height=4.3cm]{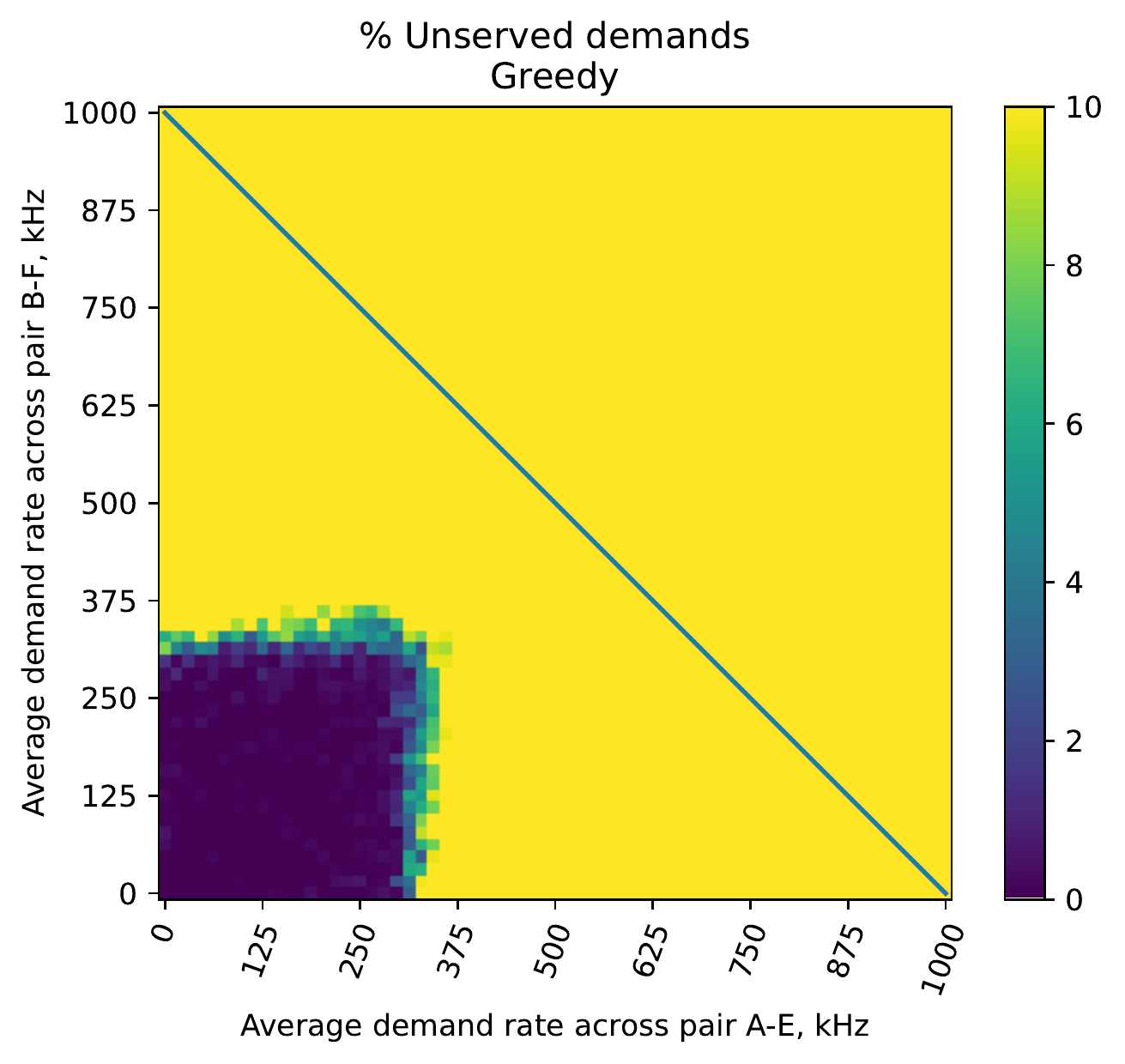}
\label{fig_resGS}}
\hfil
\subfloat[]{\includegraphics[height=4.3cm]{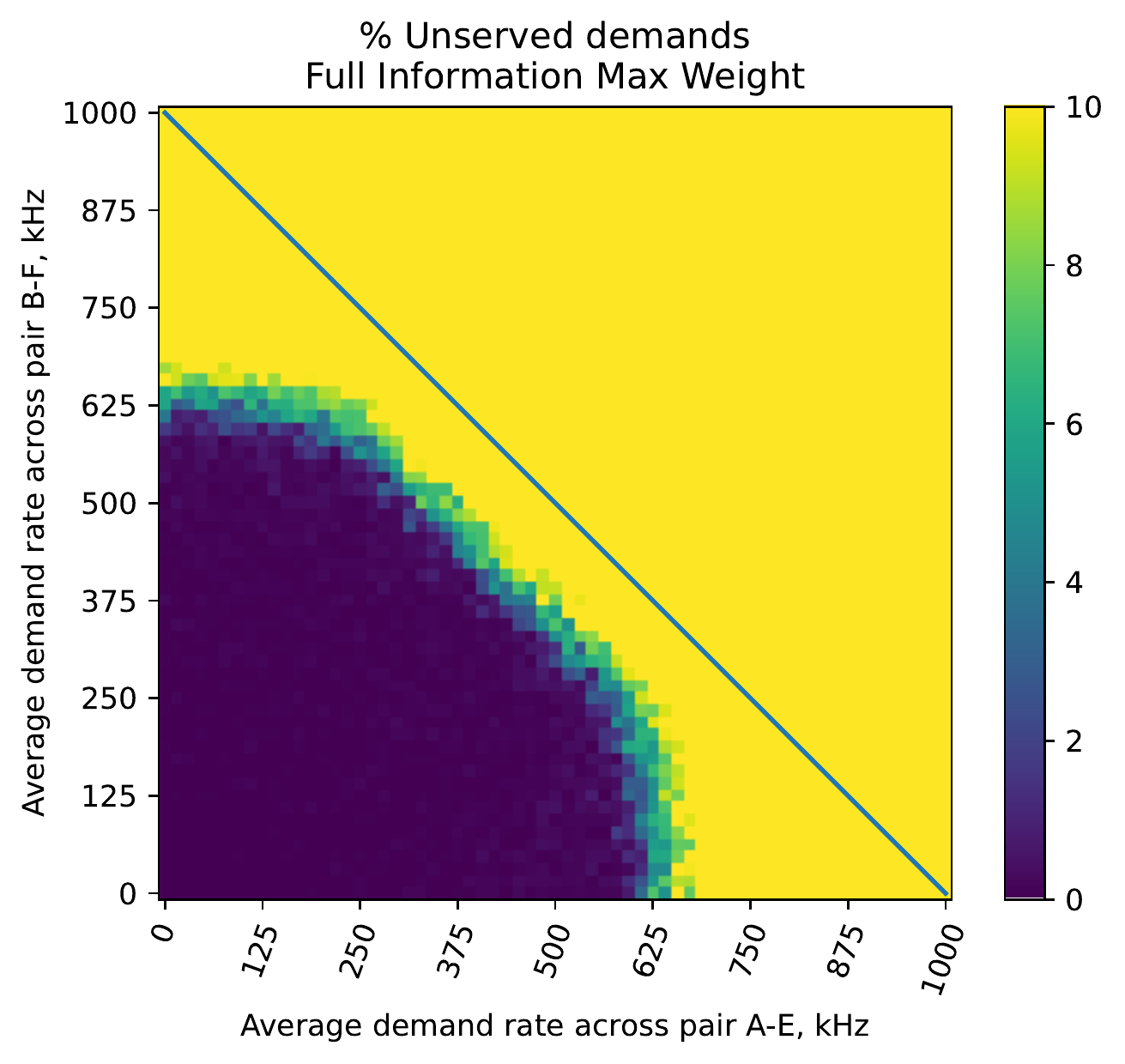}
\label{fig_resFI}}
\hfil
\subfloat[]{\includegraphics[height=4.3cm]{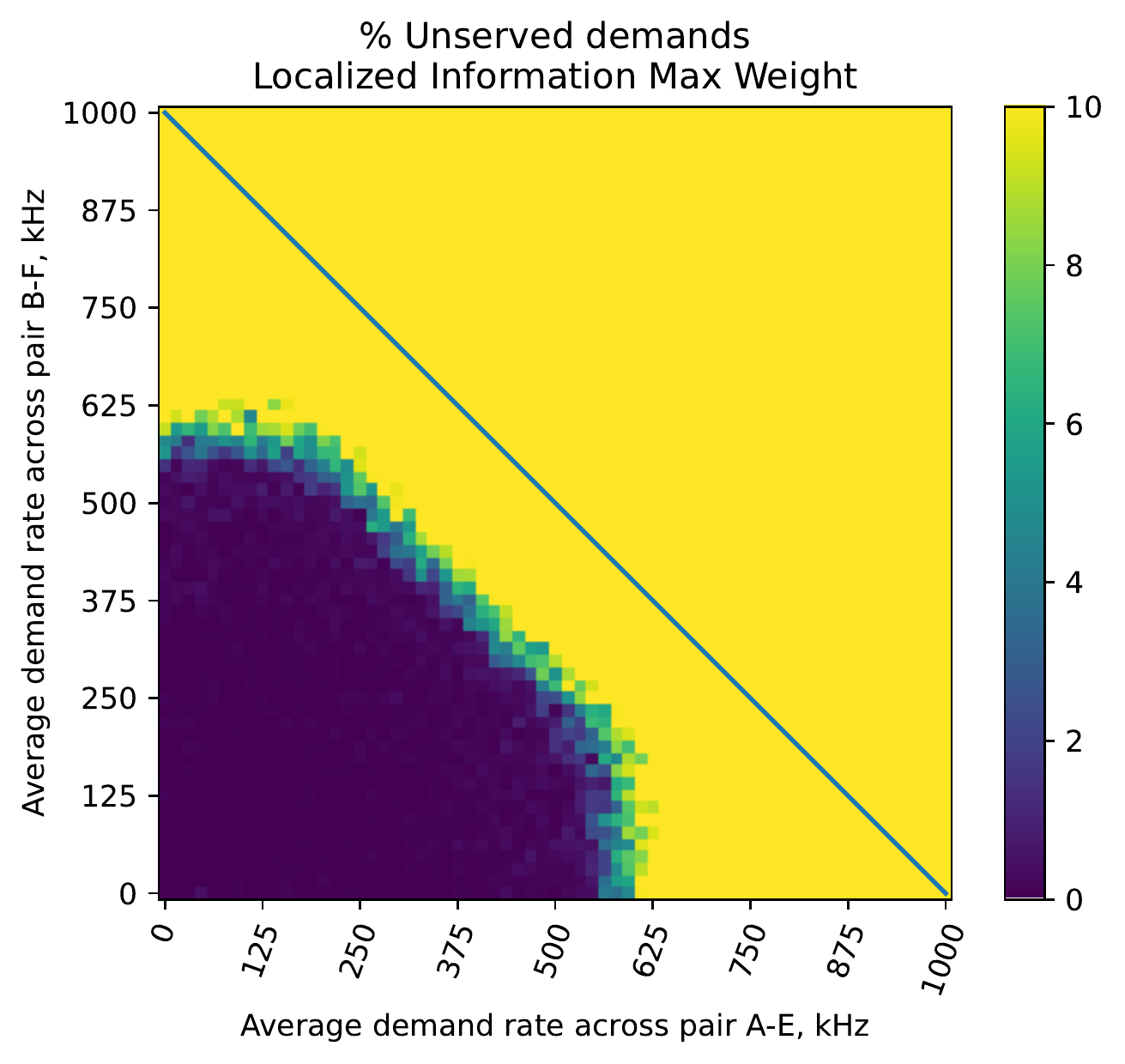}
\label{fig_resLI}}
\caption{Simulation results: percentage of unserved demands by the three schedulers. The greedy scheduler (a) exhibits a square-shaped rate region whose sides go from $0$ kHz up to $\sim 300$ kHz: this means that increasing demand across one commodity does not impair service of the other (with the current values of loss and infinite memory). The fully informed (b) and locally informed's (c) regions have a similar shape with a cut corner: the diagonal segment in the upper right corresponds to a limitation of cumulative demand, that is of the sum of the two rates. The full information can serve individual demand up to $\sim 600$ kHz, twice the performance of the greedy scheduler, and a cumulative demand of $\sim800$ kHz, while the local information can serve $\sim550$ kHz and a cumulative $\sim700$ kHz. The diagonal segment is parallel to the optimal $(\alpha,0)-(0,\alpha)$ bound (pictured), and in the lossless case it rests on it.}
\label{fig_results}
\vspace{-.5cm}
\end{figure*}
After every node $i\in\mathcal{V}$ has solved its individual problem, all the partial solutions $\mathbf{r}^i(t)$ must be blended together to create the global solution: notice that a solution that is locally feasible may not be globally so when blended with others. To solve this issue, the notion of \textit{rank} of a queue is introduced: if a queue is physical, its rank is zero. Otherwise, the rank of a queue is given by the minimum number of swapping operations required to add an ebit to it starting from an empty network. Requests are then given a random timeout, to mimic FCFS service in a real experimental system: if queues $AB$ and $BC$ are empty, then $r_{A[B]C} = 1$ means a measurement is performed and fails. This failure affects the $AC$ supply any hypothetical $A[C]i$ or $C[A]j$ transitions were scheduled to rely on, making them fail as well. On the other hand, correcting a decision for an outgoing transition for queues of rank $\geq n$ has no effect on the supply for transitions from queues of rank $< n$. Therefore, both in a real system and in our model, the users' requests are served with a FCFS discipline, in ascending rank order. The consumption part of $\mathbf{r}(t)$ must be handled with care too: since there is one term per queue and each queue is linked to two nodes, there are two possible candidates for each term. To keep a conservative approach, we take the minimum of the two candidates for each term. Notice that this is simply a design choice: other implementations could opt to take the maximum or design more complex schemes. If a conflict between a consumption operation and an intermediate scheduling operation arises, priority is always given to user service. 
We observe a great improvement in performance by this policy compared to the greedy scheduler. Moreover, as this policy is partially localized, it is practically implementable.

\section{Numerical Results}
\label{sec_numresults}
This section shows a numerical comparison among the presented scheduling policies through our algebraic model. All the results were obtained through a Python simulator that generates the $\mathbf{M}$ matrix given the Alice-Bob pairs and service routes, and evaluates different policies. Its source code is available at \cite{gitrepo}. All simulations were carried out on the $ABCDEF$ topology with $ABCDE$ and $BCDEF$ as service routes, to create a bottleneck and gauge how the different schedulers handle it. One of the standard performance metrics in classical network theory is the \emph{capacity region}, that can be described as follows: taking average demand from the competing clients as a set of axes, the capacity region is defined as the locus of points that the network can serve in a stable way, i.e., with all the queues returning to zero in a finite time. For our finite time simulation, we approximate the rate region by the percentage of unserved ebit requests (for an average user-demand), measured at the end of the simulation $t$. In Fig.~\ref{fig_results} all points in the dark blue region are certainly servable and all yellow ones are unservable, with the midway region providing an estimate for the capacity region's bound. We opt for such a performance metric (instead of the more common average queue size) for illustration purposes, since it provides us with a sharper boundary in approximating the capacity region. 
All simulations were carried out on one node of the LIP6 \texttt{small} cluster ($2$ x Intel Xeon E5645 $12$ cores, $24$ threads at $2.4$ GHz), with $10^5$ time steps of $1 \mu s$ each and an average ebit lifetime of $10 \mu s$, yielding $\eta = 0.9$. $\alpha$ was set at $1\ \text{(time steps)}^{-1} = 1$ MHz for all links. In both the fully informed and partially informed schedulers, the weight $\gamma$ of demands in the scheduling calculation was set to $1$. The greedy, full information and local information schedulers took respectively half an hour, $~ 3$ hours and $~ 10$ hours to complete their run.

Concerning the shape of the regions reported in fig. \ref{fig_results}, we remark that the ideal shape for a rate region in this context would be the $(0,0)-(\alpha,0)-(0,\alpha)$ triangle, meaning that all pairs coming to the $BCDE$ bottleneck are effectively employed in demand service.
The Full Information scheduler provides an upper bound for the potential benefits of applying Max Weight scheduling to quantum networks, while the Partial Information scheduler showcases tangible improvement with a policy that is localized enough to be reasonably implementable.

\section{Conclusions and Future Perspectives}
\label{sec_conclusion}
We have presented a novel algebraic model for scheduling in quantum networks and shown how it can be used to design original scheduling policies for arbitrary quantum network systems; these policies may be global or more localized for practicality. Aside from the simple Max-Weight policies here, other scheduling policies can be proposed and tested within our framework. Since the presented model takes static routes as inputs, an interesting extension to this work could be to integrate dynamic routing algorithms such as \cite{TowsleyGrid,routingkaushik} to obtain a full-fledged modeling toolbox to design quantum networks.
\section*{Acknowledgements}
We thank Kaushik Chakraborty for stimulating discussions. 
PF's work is funded by the French state through the \emph{Programme d’Investissements d’Avenir} managed by the Agence Nationale de la Recherche (project ANR-21-CMAQ-0001)
\bibliographystyle{IEEEtran}
\bibliography{IEEEabrv,bibliography}

\begin{thebibliography}{10}
\providecommand{\url}[1]{#1}
\csname url@samestyle\endcsname
\providecommand{\newblock}{\relax}
\providecommand{\bibinfo}[2]{#2}
\providecommand{\BIBentrySTDinterwordspacing}{\spaceskip=0pt\relax}
\providecommand{\BIBentryALTinterwordstretchfactor}{4}
\providecommand{\BIBentryALTinterwordspacing}{\spaceskip=\fontdimen2\font plus
\BIBentryALTinterwordstretchfactor\fontdimen3\font minus
  \fontdimen4\font\relax}
\providecommand{\BIBforeignlanguage}[2]{{%
\expandafter\ifx\csname l@#1\endcsname\relax
\typeout{** WARNING: IEEEtran.bst: No hyphenation pattern has been}%
\typeout{** loaded for the language `#1'. Using the pattern for}%
\typeout{** the default language instead.}%
\else
\language=\csname l@#1\endcsname
\fi
#2}}
\providecommand{\BIBdecl}{\relax}
\BIBdecl

\bibitem{1g2g3g}
\BIBentryALTinterwordspacing
S.~Muralidharan, L.~Li, J.~Kim, N.~L{\"u}tkenhaus, M.~D. Lukin, and L.~Jiang,
  ``Optimal architectures for long distance quantum communication,''
  \emph{Scientific Reports}, vol.~6, no.~1, p. 20463, Feb 2016. [Online].
  Available: \url{https://doi.org/10.1038/srep20463}
\BIBentrySTDinterwordspacing

\bibitem{TowsleySwitchProtocols}
\BIBentryALTinterwordspacing
W.~Dai, A.~Rinaldi, and D.~Towsley, ``Entanglement swapping in quantum
  switches: Protocol design and stability analysis,'' 2021. [Online].
  Available: \url{https://arxiv.org/abs/2110.04116}
\BIBentrySTDinterwordspacing

\bibitem{TowsleySwitchStochastic}
G.~Vardoyan, S.~Guha, P.~Nain, and D.~Towsley, ``On the stochastic analysis of
  a quantum entanglement distribution switch,'' \emph{IEEE Transactions on
  Quantum Engineering}, vol.~2, pp. 1--16, 2021.

\bibitem{wilde}
M.~M. Wilde, \emph{Quantum Information Theory}.\hskip 1em plus 0.5em minus
  0.4em\relax Cambridge university press, 2017.

\bibitem{protocolzoo}
\BIBentryALTinterwordspacing
``Quantum protocol zoo.'' [Online]. Available: \url{https://wiki.veriqloud.fr}
\BIBentrySTDinterwordspacing

\bibitem{RoutingMemoriesWehnerKerenidis}
\BIBentryALTinterwordspacing
E.~Schoute, L.~Mancinska, T.~Islam, I.~Kerenidis, and S.~Wehner, ``Shortcuts to
  quantum network routing,'' 2016. [Online]. Available:
  \url{https://arxiv.org/abs/1610.05238}
\BIBentrySTDinterwordspacing

\bibitem{Tassiulas}
L.~Tassiulas and A.~Ephremides, ``Stability properties of constrained queueing
  systems and scheduling policies for maximum throughput in multihop radio
  networks,'' \emph{IEEE Transactions on Automatic Control}, vol.~37, no.~12,
  pp. 1936--1948, 1992.

\bibitem{QStack}
\BIBentryALTinterwordspacing
A.~Dahlberg, M.~Skrzypczyk, T.~Coopmans, L.~Wubben, F.~Rozp\c{e}dek,
  M.~Pompili, A.~Stolk, P.~Pawe\l{}czak, R.~Knegjens, J.~de~Oliveira~Filho,
  R.~Hanson, and S.~Wehner, ``A link layer protocol for quantum networks,'' in
  \emph{Proceedings of the ACM Special Interest Group on Data Communication},
  ser. SIGCOMM '19.\hskip 1em plus 0.5em minus 0.4em\relax New York, NY, USA:
  Association for Computing Machinery, 2019, p. 159–173. [Online]. Available:
  \url{https://doi.org/10.1145/3341302.3342070}
\BIBentrySTDinterwordspacing

\bibitem{TowsleyGrid}
\BIBentryALTinterwordspacing
M.~Pant, H.~Krovi, D.~Towsley, L.~Tassiulas, L.~Jiang, P.~Basu, D.~Englund, and
  S.~Guha, ``Routing entanglement in the quantum internet,'' \emph{npj Quantum
  Information}, vol.~5, no.~1, p.~25, Mar 2019. [Online]. Available:
  \url{https://doi.org/10.1038/s41534-019-0139-x}
\BIBentrySTDinterwordspacing

\bibitem{vasantham}
\BIBentryALTinterwordspacing
T.~Vasantam and D.~Towsley, ``Stability analysis of a quantum network with
  max-weight scheduling,'' 2021. [Online]. Available:
  \url{https://arxiv.org/abs/2106.00831}
\BIBentrySTDinterwordspacing

\bibitem{DaiScheduling}
W.~Dai, T.~Peng, and M.~Z. Win, ``Optimal remote entanglement distribution,''
  \emph{IEEE Journal on Selected Areas in Communications}, vol.~38, no.~3, pp.
  540--556, 2020.

\bibitem{TowsleyScheduling}
\BIBentryALTinterwordspacing
W.~Dai and D.~Towsley, ``Entanglement swapping for repeater chains with finite
  memory sizes,'' 2021. [Online]. Available:
  \url{https://arxiv.org/abs/2111.10994}
\BIBentrySTDinterwordspacing

\bibitem{bugalho}
\BIBentryALTinterwordspacing
L.~Bugalho, B.~C. Coutinho, F.~A. Monteiro, and Y.~Omar, ``Distributing
  multipartite entanglement over noisy quantum networks,'' 2021. [Online].
  Available: \url{https://arxiv.org/abs/2103.14759}
\BIBentrySTDinterwordspacing

\bibitem{routingkaushik}
K.~Chakraborty, D.~Elkouss, B.~Rijsman, and S.~Wehner, ``Entanglement
  distribution in a quantum network: A multicommodity flow-based approach,''
  \emph{IEEE Transactions on Quantum Engineering}, vol.~1, pp. 1--21, 2020.

\bibitem{LKBMem}
\BIBentryALTinterwordspacing
M.~Cao, F.~Hoffet, S.~Qiu, A.~S. Sheremet, and J.~Laurat, ``Efficient
  reversible entanglement transfer between light and quantum memories,''
  \emph{Optica}, vol.~7, no.~10, p. 1440, oct 2020. [Online]. Available:
  \url{https://doi.org/10.1364%2Foptica.400695}
\BIBentrySTDinterwordspacing

\bibitem{gitrepo}
\BIBentryALTinterwordspacing
``Simulator github repository.'' [Online]. Available:
  \url{https://github.com/pfittipaldi/DynSchedSimulator}
\BIBentrySTDinterwordspacing

\end{thebibliography}
\end{document}